\documentclass{aastex63}

\shorttitle{Nonlinear Rossby wave-wave \& wave-mean flow theory: Long term Solar cycle modulations}
\shortauthors{Raphaldini et al.}
\usepackage{amsmath}
\usepackage{graphicx}

\begin{document}

\title{Nonlinear Rossby wave-wave and wave-mean flow theory for long term Solar cycle modulations}

%% automatically put the author and affiliation information from the manuscript
%% and save the corresponding author the trouble of entering it by hand.

\correspondingauthor{Breno Raphaldini}
\email{brenorfs@gmail.com}

\author{Breno Raphaldini}
\affil{Instituto de Astronomia, Geof\'isica e Ci\^encias Atmosfericas, Universidade de Sao Paulo}

\author{Andr\'e Seiji Teruya}

\affil{Instituto de Astronomia, Geof\'isica  e Ci\^encias Atmosfericas, Universidade de Sao Paulo}

\author{Carlos Frederico Mendonca Raupp}
\affil{Instituto de Astronomia, Geof\'isica  e Ci\^encias Atmosfericas, Universidade de Sao Paulo}

\author{Miguel D. Bustamante}
\affiliation{School of Mathematics and Statistics, University College Dublin, Belfield, Dublin 4, Ireland}

%\author{André Seiji Teruya}
%\affil{Instituto de Astronomia, Geofísica e Ciências Atmosfericas, Universidade de Sao Paulo}

\begin{abstract}
The Schwabe cycle of solar activity exhibits modulations and frequency fluctuations on slow time scales of centuries and millennia. Plausible physical explanations for the cause of these long-term variations of the solar cycle are still elusive, with possible theories including stochasticity of alpha effect and fluctuations of the differential rotation. It has been suggested recently in the literature that there exists a possible relation between the spatio-temporal structure of Solar cycle and the nonlinear dynamics of magnetohydrodynamic Rossby waves at the solar tachocline, including both wave-wave and wave-mean flow interactions. Here we extend the nonlinear theory of MHD Rossby waves presented in a previous article to take into account long term modulation effects due to a recently discovered mechanism that allows significant energy transfers throughout different wave triads: the precession resonance mechanism.  We have found a large number of Rossby-Haurwitz wave triads whose frequency mismatches are compatible with the solar cycle frequency. Consequently, by analyzing the reduced dynamics of two triads coupled by a single mode (five-wave system), we have demonstrated that in the amplitude regime in which precession resonance occurs, the energy transfer throughout the system yields significant long-term modulations on the main $\sim 11$yr period associated with intra-triad energy exchanges.  
 We further show that such modulations display an inverse relationship between the characteristic wave amplitude and the period of intra-triad energy exchanges, which is consistent with the Waldmeier's law for the solar cycle. 
In the presence of a constant forcing and dissipation, the five-wave system in the precession resonance regime exhibits irregular amplitude fluctuations with some periods resembling the Grand Minimum states.  
\end{abstract}


\section{Introduction} \label{sec:intro}

The Solar cycle observed in sunspot number time-series is approximately periodic in nature, with the main period being around 11 years. On a closer inspection, however, the solar cycle exhibits both long-term modulations and fluctuations in the 11-year period on the same time-scales. It has been suggested that several periods exist for these modulations, including 100yr, 220yr and 1000yr periods (\citealt{Usoskin2017}).

It has long been noted that an apparent connection between the duration and magnitude of the cycle does exist (\citealt{Waldmeier1936}), suggesting an inverse correlation between the amplitude of the solar activity (maximum number of sunspots at the peak of the cycle) and the duration of the cycle. Possibly, one of the most remarkable manifestations of this relationship is the period of the cycle during the historical minimum of solar activity (Maunder minimum), which was found to be increased as much as twice the usual 11-year period. Such relation has been used as an empirical prediction method of the maximal activity at the peak of the cycle as a function of the increase rate of the activity during the ascending phase of the cycle (\citealt{Pipin2012}).

All the characteristics mentioned above impose a remarkable challenge to dynamo models of practical  importance, since the  improvement of  such modelling issues may lead to better predictions of the solar activity. Although the theory behind such long-term variations of the solar cycle remains elusive, dynamo models are able to reproduce some long-term features of the solar cycle by the inclusion of a stochastically fluctuating alpha effect with slow variations. The physical basis behind such stochastic fluctuations, however, remains unclear, where one of the possible suggestions is that they are related to turbulent fluctuations in vortex sizes and their turnover time-scales (\citealt{Pipin2012}).
One particular class of models that has been able to reproduce long-term variation patterns of solar activity refers to alpha-omega dynamos with stochastic parameters (\citealt{Hoyng1993}). This class of dynamo models yields wave-like solutions of the induction equation that propagate toward the equator, therefore reproducing the butterfly pattern of solar activity, with the activity beginning at mid-latitudes, around 35 - 40 degrees of latitude, and gradually migrating toward the equator. Long-term activity variations can also be achieved in flux-transport dynamo models with a prescribed meridional flow (\citealt{Dikpati2005}).

Another feature of solar activity that is believed to exhibit fluctuations on slow time-scales is the differential rotation, or the zonal flow profile.  Although it might be difficult to detect long-term variations of the differential rotation at the tachocline or inside the convection zone, variations in differential rotation are observed at solar surface and might be associated with hemispheric asymmetries in sunspot activity (see \citealt{Zhang2015}). Such variations on the mean zonal flow are apparently in anti-correlation with amplitudes of the solar activity, suggesting a possible coupling between smaller scale processes, such as the ones that may lead to sunspot activity, and the global scale differential rotation profile.

 \cite{Raphaldini2015}  analyzed the weakly nonlinear interaction theory of MHD Rossby waves embedded in a constant toroidal magnetic field background state and showed that the periodic fluctuations of the wave amplitudes associated with the resonant triad coupling occur on the same time-scale of the solar cycle (typically around one order of magnitude greater than the linear period of the waves). Consequently, the authors suggested that the temporal energy modulations of such MHD Rossby waves due to nonlinear interaction might be related to the periodic nature of the solar magnetic activity.  

Recently, \cite{Dikipati2018} analyzed a shallow water MHD model for the solar tachocline and highlighted that the nonlinear interaction involving magnetic Rossby waves, the differential rotation profile and toroidal magnetic fields might be responsible for the so-called quasi-periodic tachocline nonlinear oscillations. Here we will argue that a similar mechanism might also give rise to long term modulations in the solar magnetic activity.

In a recent article, \cite{Bustamante2014} proposed a novel mechanism of nonlinear wave systems that might produce long term fluctuations in the wave amplitudes in an intermediate amplitude regime. The basis of this mechanism relies on a resonance between the nonlinear oscillations of one wave triad and  the fluctuations in the wave phases of an adjacent triad. In this scenario, the phases' oscillations  will be strongly influenced by the frequency mismatch among the waves. The precession resonance mechanism is shown to be responsible for strong energy transfers throughout the whole nonlinear wave system in several contexts  (\citealt{Bustamante2014}). Also, another interesting feature of this mechanism is that it allows the mean zonal flow (represented by eigenmodes having both zero time frequency and zero zonal wavenumber) to exchange energy with the waves, which is not  possible in the weak turbulence limit (small amplitude) in that the zonal flow acts as a catalyst for the energy exchanges between the waves. 

Here we augment the nonlinear interaction theory of MHD Rossby waves at the solar tachocline developed by \cite{Raphaldini2015} by accounting effects similar to the precession resonance that allows significant energy transfer throughout different triads as well as the interaction between Rossby waves and the differential rotation. For this purpose, we first search for sets of three interacting waves such that the mismatch among the waves' eigenfrequencies is close to one of the harmonics of the solar cycle, that is, $$\omega_{1}+\omega_{2}-\omega_{3} \sim j\pi/22 yr^{-1}, j = 1, 2, 3 ...  .$$ 

The triads mentioned above contain a mode with zero zonal wavenumber and zero eigenfrequency that mimics the solar differential rotation effects. Then we have analyzed a representative example of such triads in which the triad is connected via one wave mode to a second triplet. If one of these triads dominate the initial energy of the system, the initial excitation of the secondary triplet can be explained by a linearized theory through a mechanism reminiscent of modulational instability (\citealt{Connaughton2010}). However, we have demonstrated that when the nonlinearity associated  with the secondary wave triad is restored, the maximum efficiency of inter-triad energy exchange is attained in the precession resonance regime with the secondary triplet having a nonlinear frequency of amplitude modulation near $j\pi/22 yr^{-1}$, which refers to the frequency mismatch among the waves of the primary triad. The resulting energy exchanges between the two wave triads yield modulations on a time scale longer than the 22-year cycle, which corresponds to the period of the amplitude oscillations of the second triad.

In addition, as a consequence of the Manley-Rowe invariant (\citealt{Bustamante2011}), the nonlinear oscillation period of the mode amplitudes of a wave triad is inversely proportional to the square root of the energy of two modes of the triad, that is, $$T(I) \propto 1/\sqrt{I},$$ where $I$ is a weighted sum of squares of the mode amplitudes.

 The above equation provides an inverse relationship between amplitude and nonlinear period of the triad, which is remarkably similar to the aforementioned Waldmier's law. Also, the zonal flow mode amplitude modulations are found to be approximately in opposite phase with the amplitude oscillations of the second triad, which is supposedly related to the Schwabe cycle according to our theoretical model. Therefore, we argue here that the precession resonance involving MHD Rossby wave triads might be a possible mechanism behind the long-term modulations of the solar cycle observed in sunspot number time-series. 

In Section \ref{sec:style} we introduce the model equations, which refer to a simplified version of the quasi-geostrophic MHD equations derived by \cite{Zeitlin2013}, but augmented to take into account the effects of spherical geometry. Section \ref{sec:style} also revisits the linear theory of the model equations for a resting and constant toroidal magnetic field background state, as well as the reduced dynamics of a single triad of non-linearly interacting waves.  In Section \ref{sec:style} we also show that there is a considerable amount of triads of large-scale Rossby-Haurwitz modes whose frequency mismatch among the waves is comparable to the typical main frequency of the Solar cycle.  We then consider in the following sections a representative example with a set of four modes coupled to a zonal flow mode (zero eigenfrequency and zero zonal wavenumber mode). The solutions of the system are analyzed in Sections \ref{5wavedynconserv} and \ref{5wavedynforced} for the conservative and forced-dissipative cases, respectively. The main conclusions are presented in Section \ref{ConcSection}. Further details of the calculations and the mechanisms explored here are presented in the appendixes. Appendix \ref{ApendCoupCoef} presents details of the calculation of the nonlinear coupling coefficients. Appendix \ref{3WEqPRess} provides a further explanation on the two processes of excitation that are relevant to this study: the modulational instability and the aforementioned precession resonance mechanism. Appendix \ref{ApendDifus} details the evaluation of the damping coefficients.

\section{Model Equations and Wave Theory} \label{sec:style}

\subsection{Basic equations and linear theory}
We consider the barotropic vorticity equation in the rotating MHD case as described by \cite{Raphaldini2015}. This equation can be derived from the original MHD shallow water equations (\citealt{Zaqarashvili2007}, \citealt{Giman2000}) by taking the curl of the horizontal velocity and magnetic field equations, discarding the divergent terms based on dimensional arguments and therefore reducing the system to two evolution equations for the absolute vorticity and magnetic potential. Equations (1)-(2) can also be regarded as the asymptotic limit of the quasi-geostrophic MHD equations derived by  \cite{Zeitlin2013} for high equivalent depth. We assume that there is a large scale background magnetic field in the toroidal direction $\overline B(\theta)$. The evolution equations of the system in spherical coordinates are therefore written as:

 \begin{equation}
\frac{\partial q}{\partial t}+\mathcal{J}(\psi,q)=\mathcal{J}(A,\nabla^{2}A)
\end{equation}

\begin{equation}
\frac{\partial A}{\partial t}+\mathcal{J}(\psi,A)=0
\end{equation}

In the equations above, $q = \nabla^{2}\psi + 2\Omega sin\theta$ is the absolute vorticity and $A$ the magnetic potential, with $\psi$ referring to the streamfunction and $\Omega$ the rigid body rotation rate of the sun. In the spherical coordinate system adopted here the jacobian and laplacian operators take the form:

\begin{align}
\mathcal{J}(f,g)=\frac{1}{a^2cos\theta}\Big ( \frac{\partial g}{\partial \theta}\frac{\partial f}{\partial \phi}-\frac{\partial f}{\partial \theta}\frac{\partial g}{\partial \phi}\Big )  \\
\nabla^{2}f = \frac{1}{a^2cos\theta}\Big[\frac{1}{cos\theta}\frac{\partial^2 f}{\partial\phi^2} + \frac{\partial}{\partial\theta}(cos\theta \frac{\partial f}{\partial\theta})\Big]
\end{align}

for any differentiable functions $f$ and $g$, with $\phi$ being the longitude, $\theta$ the latitude and $a$ the tachocline solar radius. 
In order to linearize the equations around a background state at rest and with a zonally symmetric toroidal magnetic field in the zonal direction, $(\overline B(\theta), 0)$, we set $A=\overline A + A'$, $\psi= \psi'$ and discard the terms arising from products of perturbations. To simplify the analysis we choose $\overline A$ such that the mean toroidal magnetic field assumes the special form $\overline B=B_0cos\theta$. With the assumptions described above, the linearized perturbation equations can be written in a vector form as:

\begin{equation}\label{LinPertEq}
 \frac{\partial}{\partial t}\begin{bmatrix}\nabla^2\psi\\ A\end{bmatrix}=
\begin{bmatrix}
    -\frac{2}{a^2}\Omega\frac{\partial}{\partial \phi} & \frac{B_{0}}{a\mu_{0} \rho}\frac{\partial}{\partial\phi}\Bigg(\frac{1}{a^2}-\nabla^2\Bigg)\\
     \frac{B_{0}}{a}\frac{\partial}{\partial\phi}& 0 
  \end{bmatrix}\begin{bmatrix}\psi\\ A\end{bmatrix}
\end{equation}

In the equation above, we have omitted the primes when referring to the perturbations for simplicity. Equation (\ref{LinPertEq}) can be solved by spherical harmonics (see \citealt{Zaqarashvili2007} for similar treatise). Briefly, we assume a plane wave anzatz: 

\begin{equation}\label{LinAnsatz}
\begin{bmatrix}\psi \\ A
\end{bmatrix}=\Lambda Y_{n}^{m}(\phi, \theta)e^{-i\omega t}\overrightarrow{R} =\Lambda N_{n}^{m}P_{n}^{m}(sin\theta)e^{im\phi - i\omega t}\overrightarrow{R}
\end{equation}

with $\Lambda$ being an arbitrary constant and the associated Legendre functions satisfying the following orthogonality relation:

\begin{equation}\label{ortognlegendre}
\int_{-1}^{1}P^{m_{1}}_{n_{1}}  P^{m_{2}}_{n_{2}}dz=\frac{(n+m)!2}{(n-m)!(2n+1)}\delta_{n_1n_2}
\end{equation}

with $\delta_{n_1n_2}=1$ if $n_1=n_2$ and $0$ otherwise. The normalization constant $N_{n}^{m}$ is given by:
$$N_{n}^{m}=\bigg(\frac{(n-|m|)!(2n+1)}{(n+|m|)!}\bigg)^{\frac{1}{2}}.$$

Inserting the ansatz (\ref{LinAnsatz}) into the linearized equation (\ref{LinPertEq}) yields the following eigenvalue problem: 

\begin{equation}\label{eigenvprob}
\mathcal{L}^*\overrightarrow{R} = 	i\omega \overrightarrow{R}
\end{equation}

where the matrix $\mathcal{L}^*$ refers to the symbol of the linear operator of (\ref{LinPertEq}), that is:

\begin{equation}
 \mathcal{L}^*=\begin{bmatrix}
    -2\Omega\frac{m}{n(n+1)} & \frac{B_{0}m}{\mu_{0} \rho}\big ( \frac{1}{a^2}-\frac{2}{n(n+1)}\big)\\
   \frac{B_{0}m}{a} & 0 
  \end{bmatrix}
 \end{equation}
 
The two branches of the characteristic equation of (\ref{eigenvprob}) are defined according to

\begin{align}
\omega_{-}(m,n) = \frac{1}{2} \left(-\frac{2\Omega m}{n(n+1)} - \sqrt{\bigg(\frac{2\Omega m}{n(n+1)}\bigg)^2+\frac{4B_{0}^{2}m^{2}}{\mu_{0} \rho a^2}\Big ( 1-\frac{2}{n(n+1)}\Big)} \right) \\
\omega_{+}(m,n) = \frac{1}{2} \left(-\frac{2\Omega m}{n(n+1)} + \sqrt{\bigg(\frac{2\Omega m}{n(n+1)}\bigg)^2+\frac{4B_{0}^{2}m^{2}}{\mu_{0} \rho a^2}\Big ( 1-\frac{2}{n(n+1)}\Big)} \right)
\end{align}
  
The branch $\omega_{-}$ refers to the fast hydrodynamic mode, whilst $\omega_{+}$ represents the slow magnetic mode (\citealt{Zaqarashvili2007}). In fact, $\omega_{-}$  reduces to the classical hydrodynamic Rossby-Haurwitz wave dispersion relation for $B_0 = 0$. Note also that for $n=1$ the magnetic effects cancel out and there is only one branch corresponding to a Rossby-Haurwitz wave mode.  This agrees with the dispersion relation obtained in \cite{Zaqarashvili2007}.
The corresponding right eigenvectors $\overrightarrow{R}$ are given by

\begin{equation}\label{EigenVectorDef}
     \overrightarrow{R}_{\pm}(m,n) = 
    \begin{bmatrix}
     \frac{\omega_{\pm}(m,n)}{m}\\
     \frac{B_{0}}{a}
     \end{bmatrix}
\end{equation}

%%%%%%%%%%%%%%%%%%%%%%%%%%%%%%%%%%%%%%%%%%%%%%%%%%%%%%%%%%%%%%%%%%%

%%%%%%%%%%%%%%%%%%%%%%%%%%%%%%%%%%%%%%%%%%%%%%%%%%%%%%%%%%%%%%%%%%%

\subsection{Nonlinear theory of wave interactions}
Restoring the nonlinear terms in the perturbation approach described in the previous subsection, equations (\ref{LinPertEq}) now read: 

\begin{equation}\label{NonLinPertEq}
 \frac{\partial}{\partial t}\begin{bmatrix}\nabla^2\psi\\ A\end{bmatrix}=
 \mathcal{L}\begin{bmatrix}\psi\\ A\end{bmatrix} = \mathcal{B}\bigg(\begin{bmatrix}\psi\\ A\end{bmatrix},\begin{bmatrix}\psi\\ A\end{bmatrix}\bigg )
\end{equation}

In equation (\ref{NonLinPertEq}), the linear operator $\mathcal{L}$ is the same as in equation (\ref{LinPertEq}), and the nonlinear (bilinear) operator $\mathcal{B}$ is given by

\begin{equation}
 \mathcal{B}\bigg(\begin{bmatrix}\psi\\ A\end{bmatrix},\begin{bmatrix}\psi\\ A\end{bmatrix}\bigg) =\begin{bmatrix}
 -\mathcal{J}( \psi,\nabla^2 \psi) + \frac{1}{\mu_0\rho} \mathcal{J}( A, \nabla^2 A)\\
\mathcal{J}(\psi,  A)
 \end{bmatrix}
 \end{equation}
 
 We now consider the following ansatz in the form of a linear combination of a few number of linear eigenmodes:

\begin{equation}\label{NonLinAnsatz}
\begin{bmatrix}\psi \\ A \end{bmatrix} = \sum_{k=1}^{K}a \Lambda_{k}(t)Y_{n_{k}}^{m_{k}}(\phi,\theta) \overrightarrow{R}_{k} = \Lambda_{k}(t)N_{n_{k}}^{m_{k}}P_{n_{k}}^{m_{k}}(sin\theta)e^{im_{k}\phi}\overrightarrow{R}^{(k)}
\end{equation}

where $ \Lambda_{k }(t)$ now represents the complex valued mode amplitudes. Let us first consider the case of a single wave triplet (K = 3). In this case, we insert the ansatz (\ref{NonLinAnsatz}) into the nonlinear perturbation equations (\ref{NonLinPertEq}), and proceed to obtain the time evolution equations for the mode amplitudes $\Lambda_{k}$, $k=1,2,3$. In order to do so we make use of the orthogonality relation (\ref{ortognlegendre}), the orthogonality of the $e^{im_k \phi}$ components in the $[0, 2\pi]$ interval for different k, as well as the orthogonality of the eigenvectors $\overrightarrow{R}^{(k)}$ regarding the inner product 

\begin{equation}\label{DefInnerProd}
\langle \overrightarrow{X},\overrightarrow{Y} \rangle = X_{1}^{*}Y_{1}+\frac{1}{\mu_{0}\rho}X_{2}^{*}Y_{2}
\end{equation}
where $\overrightarrow{X}=(X_{1},X_{2})$ and $\overrightarrow{Y}=(Y_{1},Y_{2})$ represent two arbitrary elements of the $\mathbb{C}^2$ vector space, and the superscript  "*" denotes complex conjugation. Therefore, the evolution equations for the complex valued mode amplitudes are the so-called triad equations:

\begin{align}\label{TriadEqsOrig}
\frac{d\Lambda_{1}}{dt}=i\omega_{1}\Lambda_{1}+C_{1,2,3}\Lambda_{2} ^{*}\Lambda_{3} \\
\frac{d\Lambda_{2}}{dt}=i\omega_{2}\Lambda_{2}+C_{2,3,1}\Lambda_{1} ^{*}\Lambda_{3} \\
\frac{d\Lambda_{3}}{dt}=i\omega_{3}\Lambda_{3}+C_{3,1,2}\Lambda_{1} \Lambda_{2}
\end{align}

In the equations above, $C_{1,2,3}$, $C_{2,1,3}$ and $C_{3,1,2}$ are the interaction coefficients among the mode components of the triad, given by

\begin{align}\label{DefCouplingCoef}
C_{1,2,3} = \frac{-i}{2a[n_1(n_1+1)]|\overrightarrow{R}_1|^2}K^{m_{1}m_{2}m_{3}}_{n_{1}n_{2}n_{3}}\bigg[ I^{m_{1}m_{2}m_{3}}_{n_{1}n_{2}n_{3}}+L^{m_{1}m_{2}m_{3}}_{n_{1}n_{2}n_{3}}\bigg]
\end{align}

where
\begin{equation}
I^{m_{1}m_{2}m_{3}}_{n_{1}n_{2}n_{3}}= n_{2}(n_{2}+1))-(n_{3}(n_{3}+1)
\end{equation}

\begin{align}
L^{m_{1}m_{2}m_{3}}_{n_{1}n_{2}n_{3}}=\bigg(\frac{V_A}{a}\bigg)^2\Bigg( \frac{\omega_2}{m_2}-\frac{\omega_3}{m_3} \Bigg) \\
K^{m_{1}m_{2}m_{3}}_{n_{1}n_{2}n_{3}}=N_{n_{1}}^{m_{1}}N_{n_{2}}^{m_{2}}N_{n_{3}}^{m_{3}}\int_{-1}^{1}P^{m{1}}_{n_{1}}(z) \bigg( m_2 P^{m_{2}}_{n_{2}}(z)\frac{dP^{m_{3}}_{n_{3}}(z)}{dz} -m_3P^{m_{3}}_{n_{3}}(z)\frac{dP^{m_{2}}_{n_{2}}(z)}{dz} \bigg)dz
\end{align}
with $V_A =  \frac{B_0}{\sqrt{\mu_0 \rho}}$ indicating the Alfv\'en wave speed and  $z=sin\theta$. A detailed derivation of the interaction coefficients $C_{1,2,3}$, $C_{2,1,3}$ and $C_{3,1,2}$ is presented in Appendix \ref{ApendCoupCoef}. These coefficients are non zero provided the mode indexes $(m_{1},n_{1})$,$(m_{2},n_{2})$, $(m_{3},n_{3})$ satisfy the following selection rules:
\begin{align}\label{SelectRules}
m_{2}+m_{1}=m_{3} \nonumber \\
(m_{1})^2+(m_{3})^2\neq0 \nonumber \\
n_{2}n_{1}n_{3}\neq0 \nonumber \\
n_{2}+n_{1}+n_{3}\textrm{ is odd} \nonumber \\
(n^2_{2}-|m_{2}|^2)+(n^2_{1}-|m_{1}|^2) > 0 \nonumber \\
|n_{2}-n_{1}|<n_{3}<n_{2}+n_{1} \nonumber \\
(m_{2},n_{2})\neq(m_{3},n_{3}), (m_{1},n_{1})\neq(m_{3},n_{3})
\end{align}

The coupling coefficients given by (\ref{DefCouplingCoef}) can also be explicitly calculated in terms of Wigner 3j symbols (see \citealt{Jones1985} for details). To focus on the nonlinear terms only, the triad equations (\ref{TriadEqsOrig}) can  be rewritten using the change of variables  $B_k=\Lambda_k(t) exp(-i\omega_k t)$, k = 1, 2, 3, resulting in:

\begin{align}
\frac{dB_{1}}{dt}=C_{1,2,3}B_{2} ^{*}B_{3}e^{i\Delta\omega t}\label{triadmode1} \\
\frac{dB_{2}}{dt}=C_{2,1,3}B_{1} ^{*}B_{3}e^{i\Delta\omega t}\label{triadmode2} \\
\frac{dB_{3}}{dt}=C_{3,1,2}B_{1} B_{2}e^{-i\Delta\omega t}\label{triadmode3}
\end{align}

where $\Delta\omega = \omega_3 - \omega_2 - \omega_1$ is the mismatch among the triad eigenfrequencies. When $\Delta\omega=0$, the triad is said to be resonant. In the weakly nonlinear regime, in which the wave amplitudes are assumed to be small, the contribution of non-resonant triads for the nonlinear evolution of the system is usually neglected.  
The justification for this approach is essentially based on the highly truncated three-wave problem dynamics described by equations (\ref{triadmode1})-(\ref{triadmode3}). In fact, in the weakly nonlinear regime the mode amplitudes evolve in a time-scale longer than that associated with the linear wave phases. Consequently, the factor $e^{\pm i\Delta\omega t}$, in general, makes the the right hand side of (\ref{triadmode1})-(\ref{triadmode3}) highly oscillatory in time, so that the averaged contribution of the nonlinearity for the time-evolution of the wave amplitudes is rather small. The exception occurs when the triad is resonant, or nearly so, for this case the nonlinearity yields significant energy exchanges among the triad components. The predominance of resonant triads in the weakly nonlinear regime can also be demonstrated by near identity transformation (\citealt{Zakharov2012}). \cite{Raphaldini2015} analyzed the solutions of equations (\ref{triadmode1})-(\ref{triadmode3}) for resonant triads of MHD Rossby modes. They showed that in the amplitude regime in which all the three modes undergo significant energy modulations, the nonlinear amplitude oscillations have a period compatible with the Schwabe cycle. Consequently, they argued that these temporal energy modulations of MHD Rossby waves due to resonant triad interaction might be related to the periodic nature of the solar magnetic activity.  

 Nevertheless, for a system of several connected wave triplets, \cite{Bustamante2014} demonstrated that even if the linear frequency mismatch $\Delta\omega$ is large, strong energy transfer can occur between different wave triads provided the linear frequency mismatch of a wave triad resonates with the characteristic nonlinear frequency of the system (for example, the frequency of energy exchange associated with an adjacent wave triad). This novel mechanism of nonlinear wave systems is called precession resonance and occurs in an energy level that is not sufficiently small to neglect non-resonant wave triplets, but is still small enough to make the linear frequency mismatch $\Delta\omega$ to dominate the time evolution of the combination of the phases of the complex valued mode amplitudes $\Lambda_j(t)$, j = 1, 2, 3. Consequently, it was shown that this mechanism is able to promote events of strong energy exchanges between waves even if they are non-resonant. A more detailed description of the precession resonance mechanism for a system of two connected wave triplets is presented in Appendix \ref{3WEqPRess}.

\begin{figure}
  \includegraphics[width=\linewidth]{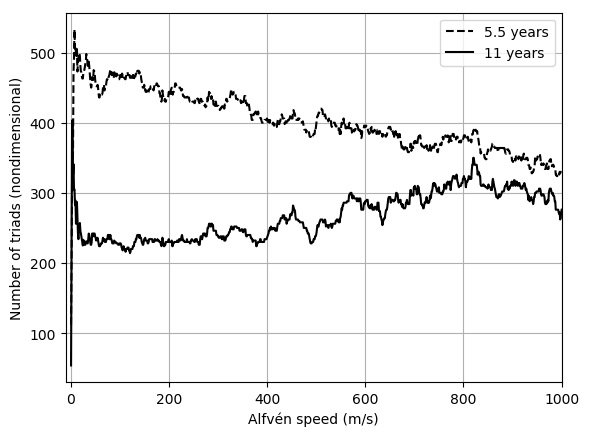}
  \caption{Number of triads whose mismatch among the mode eigenfrequencies corresponds to a period of  5.5 or 11 years (with 10 \% of tolerance), as a function of the Alfv\'en wave speed $V_A = \frac{B_0}{\mu_0 \rho}$. The periods of 5.5 and 11 years correspond to resonances of the type 4:1 and 2:1, respectively, with the Schwabe cycle. We consider only triads involving a zonal mode with spherical harmonic degree 1, 2, 3 or 4. The search has been truncated for harmonics with degree and order up to 30. We observe that the number of triads, of the order of hundreds, satisfying the above conditions, is abundant for any value of Alfv\'en wave speed up to 1000 m/s.}
  \label{FigNumbertriads}
\end{figure}

Therefore, in order to investigate the potential role of the precession resonance mechanism in promoting significant energy transfer between different wave triads in our MHD Rossby wave context, as well as to analyze its potential role in yielding long-term modulations in the Schwabe cycle of solar magnetic activity, we have sought triads whose linear frequency mismatch among the modes is close to one of the harmonics of the main solar cycle frequency, that is $\Delta\omega \approx j\pi/22  yr^{-1}, j = 1, 2, 3 ... $ . Another condition imposed for the intended triads is that they contain a mode with zero zonal wavenumber (and, consequently, zero eigenfrequency) that mimics the zonal flow effects of solar differential rotation. This condition is based on the observational results of \cite{Zhang2015}, who showed that modulations of the Schwabe cycle exhibit significant negative correlation with observed variations of the solar differential rotation strength. The number of such wave triplets as a function of the Alfv\'en wave speed is displayed in Fig. \ref{FigNumbertriads} for resonances 2:1 ($\Delta\omega \approx 4\pi/22 yr^{-1}$) and 4:1 ($\Delta\omega \approx 8\pi/22 yr^{-1}$).

In the next section we shall analyze the reduced dynamics in which a representative example of the triads mentioned above is connected with a nearly resonant triad whose nonlinear interaction period is close to the Schwabe cycle period, which in turn is one order of magnitude longer than the linear period of the waves. As we will demonstrate, strong energy transfer between the two wave triads occurs due to: a modulational type of instability and the precession resonance mechanism mentioned above, which enhances the efficiency of energy transfer between two adjacent wave triplets in the unstable regime. This inter-triad energy exchange yields significant longer period modulations on the intra-triad energy exchanges that exhibit a period compatible with long time modulations of the Schwabe cycle.

\section{Nonlinear 5-wave model in the conservative case}\label{5wavedynconserv}
Let us consider now the ansatz (\ref{NonLinAnsatz}) for K = 5, in which modes 1, 2 and 3 satisfy the conditions (\ref{SelectRules}), and modes 3, 4 and 5 satisfy the same selection rules, apart from the resonance condition for their eigenfrequencies ($\omega_3 \approx \omega_4 + \omega_5$). In this way, substituting the ansatz (\ref{SelectRules}) into the nonlinear perturbation equations (\ref{NonLinPertEq}) yields:

\begin{align}
\frac{d\Lambda_{1}}{dt}=i\omega_{1}\Lambda_{1}+C_{1,2,3}\Lambda_{2} ^{*}\Lambda_{3} \label{FiveWaveEquationsa} \\
\frac{d\Lambda_{2}}{dt}=i\omega_{2}\Lambda_{2}+C_{2,3,1}\Lambda_{1} ^{*}\Lambda_{3} \label{FiveWaveEquationsb} \\
\frac{d\Lambda_{3}}{dt}=i\omega_{3}\Lambda_{3}+C_{3,1,2}\Lambda_{1} \Lambda_{2} + C_{3,4,5}\Lambda_{4}^* \Lambda_{5} \label{FiveWaveEquationsc} \\
\frac{d\Lambda_{4}}{dt}=i\omega_{4}\Lambda_{4}+C_{4,3,5}\Lambda_{5} \Lambda_{3}^* \label{FiveWaveEquationsd} \\
\frac{d\Lambda_{5}}{dt}=i\omega_{5}\Lambda_{5}+C_{5,3,4}\Lambda_{4} \Lambda_{3} \label{FiveWaveEquationse}
\end{align}

Equations (\ref{FiveWaveEquationsa})-(\ref{FiveWaveEquationse}) have been integrated numerically by using an explicit eighth-order Runge-Kutta discretization. A representative example of two wave triplets constituting the five wave model (\ref{FiveWaveEquationsa})-(\ref{FiveWaveEquationse}) is displayed in Table 1. The triplets are labeled by \textbf{a} (Modes 1, 2 and 3) and \textbf{b} (Modes 3, 4 and 5), and are composed of magnetic-branch (slow) modes having spherical harmonics (0,2), (1, 10), (1, 9) and (1, 9), (1, 12), (2, 10), respectively. Recall that Triad \textbf{a} has a frequency mismatch of around 5.5 years, whereas Triad \textbf{b} is a nearly resonant wave triplet with a mismatch among the mode eigenfrequencies of the order of 15 years. 

\begin{figure}
  \includegraphics[width=\linewidth]{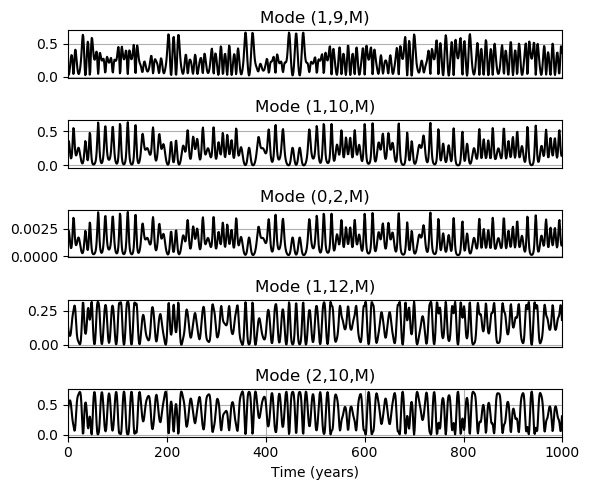}
  \caption{Time evolution of the mode energies in the conservative  5-wave model.  The modes 1, 2, 3, 4 and 5 are characterized, respectively, by the spherical harmonics (0, 2) , (1, 10) ,(1, 9), (1, 12) and (2, 10), all in the slow branch.}
  \label{Figure1}
\end{figure}

Fig. \ref{Figure1} displays the result of a numerical integration of the five-wave model (\ref{FiveWaveEquationsa})-(\ref{FiveWaveEquationse}) for the representative example illustrated in Table 1. In this integration, the initial mode amplitudes of Triad \textbf{b} are set to match the precession resonance regime, in which the characteristic frequency of amplitude modulation of this triad exhibits a 2:1 resonance with the eigenfrequency mismatch of Triad \textbf{a}. From the time evolution of the mode energies presented in Fig. \ref{Figure1}, one can notice the shortest period of energy oscillation of the order of 10 years, which is associated with the energy exchanges within each triad. This refers to the half of the period of the corresponding mode amplitude oscillation. Apart from the main $\approx 10yr$ period associated with intra-triad coupling, the time evolution of the mode energies displayed in Fig. \ref{Figure1} exhibits the alternation of small and large peaks, indicating that the magnitude of the $\approx 10yr$ cycle is modulated on a longer time-scale. Comparing the time evolution of the energies of modes (0,2) and (1, 12), which pertain to different wave triads, shows that their large and small energy peaks are approximately in opposite phase with each other. This points out that the longer time-scale modulation of the main $\approx 10yr$ cycle is a result of the inter-triad energy transfers. Therefore, Fig. \ref{Figure1} shows that, in the precession resonance regime, a strong energy transfer  between the adjacent wave triads takes place, yielding long-term modulations on the main $\approx 10yr$ cycle associated with intra-triad energy exchanges. 

There are two processes of excitation that are relevant to this study. The first process is based on \textbf{instability} and considers small initial amplitudes for the modes 1 and 2, pertaining to triad \textbf{a}. The process is reminiscent of modulational instability (\citealt{Connaughton2010}). By linearizing the system around a quasi periodic solution of the isolated  triad \textbf{b} (modes 3, 4 and 5) we calculate the maximal Lyapunov exponent of the full set of equations, yielding a growth rate of 0.29/yr (see Appendix \ref{3WEqPRess}). The second process is fully nonlinear and is based on \textbf{precession resonance} (\cite{Bustamante2014}). There, as the system explores several amplitude levels due to forcing and dissipation, at certain mode amplitudes a balance can be struck between a linear combination of the  frequency mismatches of the two triads and the nonlinear frequency broadening stemming from the finiteness of the amplitudes. At such amplitudes, a low-frequency oscillation is generated which can lead to strong energy transfers across modes. When some of the triads are quasi-resonant, the amplitudes at which precession resonance occurs can be quite small and therefore attainable in real situations. In Appendix \ref{3WEqPRess} we study briefly this mechanism for system (\ref{FiveWaveEquationsa})-(\ref{FiveWaveEquationse}), yielding energy transfer efficiencies of up to 34\%.

\begin{figure}
  \includegraphics[width=\linewidth]{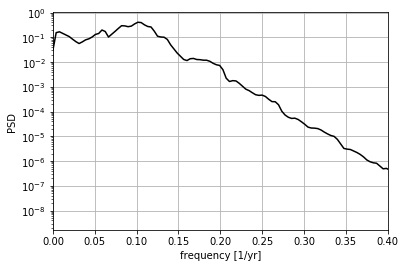}
 \caption{Power spectral density of Mode (1,9) energy time-series referred to the conservative 5-wave model integration displayed in Fig. \ref{Figure1}. There is one primary peak in the spectrum at the period of 10years, and a secondary peak associated with a modulation with a period around 120 years.}
  \label{Figure2}
\end{figure}

In order to better quantify the periods involved in the time evolution of the mode energies presented in Fig. \ref{Figure1}, Fig. \ref{Figure2} shows the power spectrum referred to the time series corresponding to the energy of Mode (1, 9) (Mode 3). The spectrum has been calculated by using  Welch's method (\citealt{Welch1967}). From Fig. \ref{Figure2} one observes a main peak at around the 10 year period, and a secondary peak at around 130 years,  which is reasonably compatible with the Gleissberg cycle (\citealt{Usoskin2017}). 

\begin{figure}
  \includegraphics[width=\linewidth]{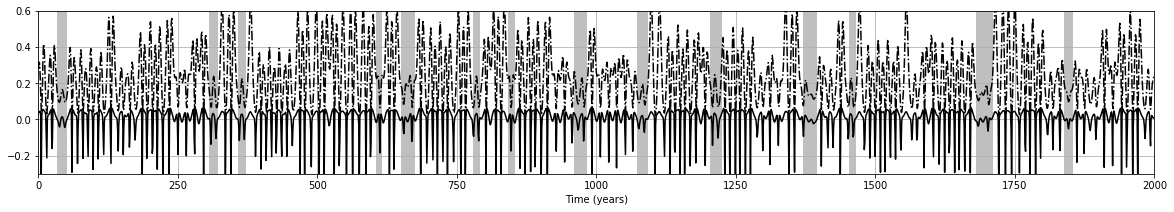}
 \caption{Instantaneous frequency of mode (1,9)  referred to the same numerical integration displayed in Figs. \ref{Figure1} and \ref{Figure2}.}
 \label{Figure3}
\end{figure}

Recall that in the case of a single wave triplet, whose dynamics is described by the three-wave equations (\ref{triadmode1})-(\ref{triadmode3}), the time evolution of the wave amplitudes (energies) is exactly periodic in time, with the solutions being described in terms of Jacobi elliptic functions, as described in Appendix \ref{3WEqPRess} (see also \citealt {Raphaldini2015} and references therein). In addition, as a consequence of the Manley-Rowe invariants, the nonlinear oscillation period of the mode amplitudes of a wave triad (Triad \textbf{b}, for instance) is inversely proportional to the square root of the energy of two modes of the triad, that is, $$T(I) \propto 1/\sqrt{I},$$ where $I=|\Lambda_{3}|^2+|\Lambda_{4}|^2$ is the Manley-Rowe invariant (see \citealt{Bustamante2011} for details). However, in the five-wave problem described by (\ref{FiveWaveEquationsa})-(\ref{FiveWaveEquationse}), the quantity $I$ of a wave triad becomes variable in time due to its coupling to the adjacent wave triplet, and so does the period of triad amplitude oscillation $T(I)$. Consequently, in the five-wave model dynamics the characteristic interaction period $T$ of a wave triad is decreased (increased) during the periods of large (small) energy peaks of the correspondent wave triad. As the characteristic interaction period $T$ of Triad \textbf{b} is compatible to the time-scale of the Schwabe cycle of solar magnetic activity, this relationship between the interaction period of a wave triad and the correspondent triad energy level is consistent with the well-known Waldmier's law for the solar cycle. To verify this relation we have computed the instantaneous frequency of the spectral amplitudes $\Lambda_k(t)$ by applying the Hilbert transform. Fig. \ref{Figure3} shows the time evolution of the instantaneous frequency of the amplitude of Mode (1,9), which shows that the frequency increases (decreases) during periods of high (low) amplitudes of the $10 yr$ cycle.

\section{Nonlinear 5-Wave Model with Forcing and Dissipation}\label{5wavedynforced}
As waves in the solar tachocline are subjected to the action of forcing and dissipation, here we investigate how these effects can modify the 5-wave dynamics in the precession resonance regime. As argued in \cite{Raphaldini2015}, the forcing acting on barotropic Rossby waves comes from the horizontal divergence of both 2-dimensional velocity and magnetic fields. This approach of considering a prescribed zero-mean horizontal divergence field as a Rossby wave source is usual  in studies of Rossby waves in Earth's Atmosphere (\citealt{Hoskins1981} and references therein). In the solar tachocline, the horizontal divergence of the velocity field stems from different physical processes such as baroclinic instability (\citealt{Gilman2014}), gravity waves and nonhomogeneus thermal forcings at the top of the radiative zone. In this context, the generalization of the nonlinear perturbation equations (\ref{NonLinPertEq}) for the forced-dissipative case is given by

\begin{equation}\label{ForcedNonLinPertEq}
 \frac{\partial}{\partial t}\begin{bmatrix}\nabla^2\psi\\ A\end{bmatrix}=
 \mathcal{L}\begin{bmatrix}\psi\\ A\end{bmatrix} = \mathcal{B}\bigg(\begin{bmatrix}\psi\\ A\end{bmatrix},\begin{bmatrix}\psi\\ A\end{bmatrix}\bigg ) + \mathbf{F} + \mathcal{D}\bigg(\begin{bmatrix}\psi\\ A\end{bmatrix},\begin{bmatrix}\psi\\ A\end{bmatrix}\bigg )
\end{equation}

where the prescribed forcing vector $\mathbf{F}$ and the damping operator $\mathcal{D}$ are 

\begin{equation}
 \mathbf{F} = \bigg[\begin{matrix}-2\Omega   sin\theta   D_u(\phi, \theta, t)\\ 0\end{matrix}\bigg ] \\
\end{equation} 
 
 \begin{equation}\label{DifusionOperator}
\mathcal{D} =
\begin{bmatrix}\nu\nabla^2(\nabla^2) & 0\\  0& \eta  \nabla^2(\nabla^2) \end{bmatrix} 
 \end{equation}

with $D_u(\phi, \theta, t)$ indicating the prescribed horizontal divergence of the velocity field. We have omitted the nonlinear terms involving $D_u$, as well as the nonlinear terms involving the divergence of the 2-dimensional magnetic field on the right hand side of the magnetic potential equation, assuming them to be small. In (\ref{DifusionOperator}), the parameters $\nu$ and $ \eta$ are the coefficients of viscous and magnetic diffusivity, respectively. Here we have utilized the values of $\nu=2.7 \times 10 cm^2 /s$ and  $\eta=4.1 \times 10^2 cm^2 /s$, as suggested in \cite{Gough2007}. We have further assumed that the horizontal divergence field (actually, $2\Omega sin\theta D_u$) has the same spatial structure of Mode 3, which is the mode that couples the two triads, and the time dependence of the forcing resonates with this mode. The effect of these assumptions is to yield a constant forcing coefficient only on the RHS of the time evolution equation of Mode 3 amplitude. %Since this mode is the unstable mode of both wave triplets, this condition also guarantees that the total energy of the system remains bounded in time irrespective of the damping (see \citealt {Harris2012} and \citealt{Raupp2009}). 

With the above considerations, the resulting generalization of the 5-wave equations (\ref{FiveWaveEquationsa})-(\ref{FiveWaveEquationse}) with the inclusion of the forcing and damping is

\begin{align}
\frac{d\Lambda_{1}}{dt}=i\omega_{1}\Lambda_{1}+C_{1,2,3}\Lambda_{2} ^{*}\Lambda_{3} - d_1\Lambda_1 \label{FiveWaveEquationsFa} \\
\frac{d\Lambda_{2}}{dt}=i\omega_{2}\Lambda_{2}+C_{2,3,1}\Lambda_{1} ^{*}\Lambda_{3} - d_2\Lambda_2\label{FiveWaveEquationsFb} \\
\frac{d\Lambda_{3}}{dt}=i\omega_{3}\Lambda_{3}+ f_3 + C_{3,1,2}\Lambda_{1} \Lambda_{2} + C_{3,4,5}\Lambda_{4}^* \Lambda_{5} - d_3\Lambda_3 \label{FiveWaveEquationsFc} \\
\frac{d\Lambda_{4}}{dt}=i\omega_{4}\Lambda_{4}
+C_{4,3,5}\Lambda_{5} \Lambda_{3}^* - d_4\Lambda_4\label{FiveWaveEquationsFd} \\
\frac{d\Lambda_{5}}{dt}=i\omega_{5}\Lambda_{5}+C_{5,3,4}\Lambda_{4}\Lambda_{3} - d_5\Lambda_5\label{FiveWaveEquationsFe}
\end{align}

where 

\begin{equation}\label{FinalExpressCoefDifus}
d_{i}=
\Bigg\langle \overrightarrow{R}^{(i)},
\begin{bmatrix}
   \nu(n_i(n_i+1)) & 0\\
     0& \eta  (n_i(n_i+1))
  \end{bmatrix}
 \overrightarrow{R}^{(i)} \Bigg\rangle
\end{equation}

for i = 1, 2, 3, 4, 5, and the coefficient $f_3$ is a constant that depends on the magnitude of the divergence forcing. A more thorough derivation of the damping coefficients is presented in Appendix \ref{ApendDifus}.

\begin{figure}
  \includegraphics[width=\linewidth]{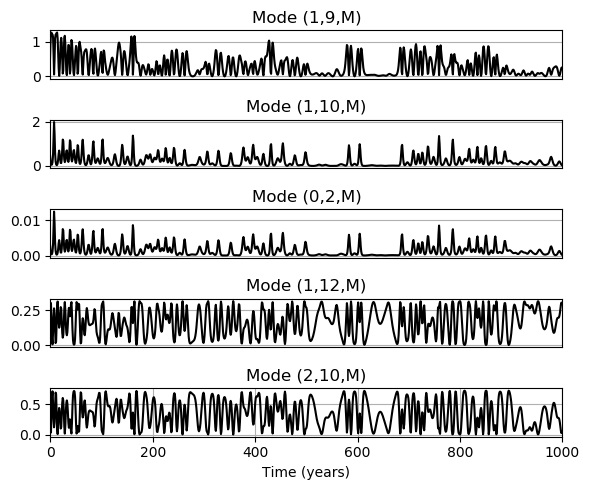}
  \caption{Similar to Fig. \ref{Figure2}, but for the forced-damped case. We observe in certain periods of the time series of the first three wave modes that they synchronize in a "Maunder-minimum-like" behavior.}
  \label{Figure4}
\end{figure}

\begin{figure}
  \includegraphics[width=\linewidth]{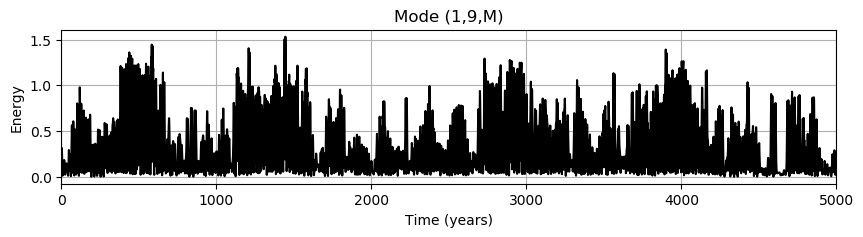}
  \caption{Same as Fig. \ref{Figure4}, but for a longer time-integration.}
  \label{Figure5}
\end{figure}

\begin{figure}
  \includegraphics[width=\linewidth]{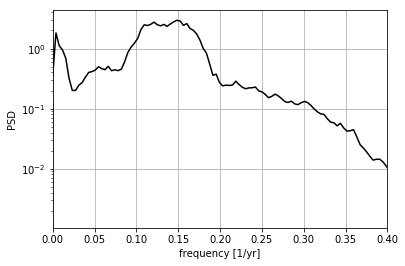}
  \caption{Power spectral density of Mode (1,9) energy time-series referred to the forced-damped 5-wave model integration displayed in Fig. \ref{Figure4}. There is one primary broad peak in the spectrum with a period of 7.5 - 9 years, and a secondary peak that gives the modulation with a period around 250 years.}
  \label{Figure6}
\end{figure}

 Results of the numerical integration of system (\ref{FiveWaveEquationsFa})-(\ref{FiveWaveEquationsFe}) are presented in Figs. \ref{Figure4} and \ref{Figure5}. Fig. \ref{Figure4} displays the time evolution of the mode energies during a 1000-year period, whereas Fig. \ref{Figure5} shows the time evolution of Mode 3 energy correspondent to the same numerical solution of Fig. \ref{Figure4}, but for a longer period of integration (5000 years). As in the conservative case, from Fig. \ref{Figure4} one notices oscillations in the mode energies on a decadal time-scale superposed by modulations of the energy peaks on a time-scale of centuries.   Again, to better quantify the main oscillation periods involved in the time evolution of the mode energies, Fig. \ref{Figure6} shows the power  spectrum computed from the Mode 3 energy time-series (i.e., the mode with spherical harmonic (1, 9)). In comparison with the spectrum obtained in the conservative system, one can notice in the forced-damped case a broader main spectral peak, which is also slightly shifted to a period of 7-9 years. Also, apart from this broad spectral peak band around 8 years, Fig. \ref{Figure6} shows a spectral peak corresponding to long term modulations with period around 230 years. 

One remarkable feature on the 5-wave model with forcing and dissipation is that the evolution of the mode energies presents some periods of suppressed activity that resemble the Maunder Minimum. This fact can be more clearly illustrated in the longer time integration presented in Fig. \ref{Figure5}. One can observe in Fig. \ref{Figure5} the appearance of several periods with very low activity lasting several decades.  Other integrations with different values of the forcing parameter $f_3$ show that the duration of such periods is highly dependent on the magnitude of the divergence forcing (figures not shown).

\section{Conclusions}\label{ConcSection}
Here we have augmented the nonlinear interaction theory of MHD Rossby waves in the solar tachocline developed by \cite{Raphaldini2015} to take into account the effect of the precession resonance mechanism that allows significant energy transfer throughout different wave triads as well as the interaction between Rossby waves and modes having zero zonal wavenumber and zero eigenfrequency, which are believed to contribute to the zonal flow profile associated with the solar differential rotation. For this purpose, we have sought interacting triads containing a zero zonal wavenumber mode yielding unstable solutions (in the modulational sense according to \citealt{Connaughton2010} ).

Consequently, we have analyzed a representative example of such triads in which the triad is connected via one wave mode to a second triplet that is nearly resonant. Numerical integrations of the five wave system show that the energy transfers between the two wave triplets allowed by modulational type instability yield long-term modulations on the main approximately 11/22 year cycle associated with intra-triad energy exchanges. In addition, the zonal flow mode amplitude modulations are found to be approximately in opposite phase with the amplitude oscillations of the second triad, which is supposedly related to the Schwabe cycle according to our theoretical model. This result is consistent with the observational work of \cite{Zhang2015}, who showed that modulations of the Schwabe cycle exhibit significant negative correlation with observed variations of the solar differential rotation strength. 

 When analyzing the dynamics of the five-wave system in the presence of a divergence forcing and dissipation, a remarkable resemblance is found between the time evolution of the wave amplitudes and the observed long-term variations of the solar cycle, with a 11/22 year cycle being modulated at time scales one order of magnitude longer (~100 years), as well as the emergence of periods of suppressed wave activity lasting several decades that resemblance the Grand minimum states. In addition, we have demonstrated that the amplitude of the Rossby wave “activity” is inversely proportional to the instantaneous period of nonlinear energy exchange. Similarly, observations of the solar cycle point out that the amplitude of the cycle, which is commonly measured by the number of sunspots at the peak phase of the cycle, is inversely proportional to the duration of the cycle. This relation between the strength and duration of the solar cycle is described by the so-called Waldmier's law. Therefore, we argue here that the modulational-like instabilities involving MHD Rossby wave triads might be a possible mechanism behind the long-term modulations of the solar cycle observed in sunspot number time-series.

It was shown recently by \cite{Raphaldini2015}  that the propagation of the magnetic branch of MHD Rossby modes is confined to an equatorial belt extending from -35$^{\circ}$ to +35$^{\circ}$ in latitude and refracted towards the equator, similar to the sunspot evolution during the solar cycle depicted by the butterfly diagram. We believe that combining the arguments of the present paper  with \cite{Raphaldini2015}  makes MHD Rossby waves strong candidates to play a major role in the dynamics of the solar activity. We have therefore to speculate the possible link between Rossby waves and the solar magnetic activity.

A dynamo model provided by Rossby wave motions was first suggested by \cite{Gilman1969a} and \cite{Gilman1969b}, by using a two-layer quasi-geostrophic model. As further argued recently by the same author \cite{Gilman2014}, baroclinic Rossby waves and baroclinic instability should be able to account for a dynamo mechanism, since they provide both vorticity and small vertical motions, which constitute necessary physical ingredients for the alpha effect, which is known to be associated with helicity. Such combination of vorticity and vertical motions can also be able to amplify the poloidal component of the magnetic field at the expense of the toroidal one. Smaller scale instabilities could also create ascending filaments of magnetic field associated with sunspots. 
Possible extensions of the present study include the analysis of larger clusters of nonlinearly interacting triads and the possibility of emergence of self-organized synchronized states, such as in \citealt{Chian2010} that could explain the approximately periodic nature of the solar dynamo.

\section*{Acknowledgement} B. Raphaldini would like to thank FAPESP (process 17/23417-5) and the FAPESP-PACMEDY project (process 15/50686-1). The work of Carlos F. M. Raupp and Andr\'e Teruya was supported by CAPES IAG/USP PROEX (process 0531/2017). Carlos Raupp also acknowledges the support from FAPESP-PACMEDY project (process 15/50686-1).

\appendix

\section{Coupling coefficients}\label{ApendCoupCoef}
In Section 2 we introduced the coupling coefficients that arise from the nonlinear terms in the perturbation equations, which have the form of a Jacobian operator $\mathcal{J}(.,.)$. Here we provide a more detailed description of the derivation of such coefficients.
The definition of the interaction coefficient $C_{j, k, l}$ involving three arbitrary modes $j, k$ and $l$ is the projection, in terms of pseudoenergy norm, of the nonlinear term applied to modes $l, k$ onto the first mode $j$:
\begin{equation}
C_{j,k,l}=\frac{1}{E_j}\int_0^{2\pi} \int_{-\frac{\pi}{2}}^{\frac{\pi}{2}}\langle
\mathcal{B}(\mathbf{u_{k}},\mathbf{u_{l}})+\mathcal{B}(\mathbf{u_{l}},\mathbf{u_{k}}),\mathbf{u_{j}}\rangle a^2 cos\theta d\theta d\phi
\end{equation}
where the vector $\mathbf{u}_{k}$ describes the spatial structure of a particular eigensolution of the linear perturbation equations (similarly for the $j$-th and $l$-th modes):
\begin{equation}
\mathbf{u_{k}}(\phi, \theta)= 
\begin{bmatrix}
\psi_k(\phi, \theta)\\
A_k(\phi, \theta)
\end{bmatrix}
\end{equation}

with
\begin{equation}\label{AnsatzApendice1}
\psi_k(\phi, \theta) = N_{n_{k}}^{m_{k}}P_{n_{k}}^{m_{k}}(\sin\theta)e^{im_{k}\phi}R_1^{(k)}
\end{equation}

\begin{equation}\label{AnsatzApendice2}
A_k(\phi, \theta) = N_{n_{k}}^{m_{k}}P_{n_{k}}^{m_{k}}(\sin\theta)e^{im_{k}\phi}R_2^{(k)}
\end{equation}

In the equations above, $R_1^{(k)} = \frac{\omega_k}{m_k}$ and $R_2^{(k)} = \frac{B_0}{a}$ refer to the components of the corresponding eigenvector $\overrightarrow{R}^{(k)}$ defined by (\ref{EigenVectorDef}) and $E_j$ is the pseudoenergy norm of the $j$-th eigenmode, given by

\begin{equation}
E_j= \int_0^{2\pi} \int_{-\frac{\pi}{2}}^{\frac{\pi}{2}}\bigg[|\nabla\psi_j|^2 + \frac{1}{\mu_0\rho}|\nabla A_j|^2 \bigg] a^2 cos\theta d\theta d\phi
\end{equation}

Therefore, to evaluate the nonlinear coupling coefficient $C_{j, k, l}$, one needs to obtain the pseudoenergy norm from the nonlinear perturbation equations (\ref{NonLinPertEq}). As the first equation is written for $\nabla^2 \psi$ and the second is written for $A$, one must multiply the first equation by $\psi^*$ and the second by $\nabla^2 A^*$ and integrate by parts to yield the pseudoenergy norm, where the superscript * denotes the complex conjugate. Consequently, to be consistent with the pseudoenergy norm one projects the nonlinear term onto  the corresponding adjoint eigensolution  $\mathbf{u}_{j}^{\dagger}$ referred to the $j$-th mode, given by

\begin{equation}
\mathbf{u_{j}}^{\dagger}=
\begin{bmatrix}
 \psi_j^*\\
\nabla^2 A_j^*
\end{bmatrix}
\end{equation}

In this way, substituting the ansatz (\ref{NonLinAnsatz}) into the nonlinear perturbation equation (\ref{NonLinPertEq}), multiplying the resulting equations by the adjoint eigensolution of mode $j$ given above, integrating by parts the resulting equations and using the boundary conditions (periodic solutions in $\phi$ and regularity at the poles), as well as the orthogonality relations, we get:

\begin{equation}\label{EqSpecAmplitGenericApendTriad}
E_j \frac{d\Lambda_j}{dt} -E_j i\omega_j \Lambda_j = \Lambda_k(t) \Lambda_l(t) \int_0^{2\pi} \int_{-\frac{\pi}{2}}^{\frac{\pi}{2}}\langle
\mathcal{B}(\mathbf{u_{k}},\mathbf{u_{l}})+\mathcal{B}(\mathbf{u_{l}},\mathbf{u_{k}}),\mathbf{u_{j}}\rangle a^2 cos\theta d\theta d\phi
\end{equation}

where the nonlinear operator is given by

\begin{equation}
 \mathcal{B}(\mathbf{u}, \mathbf{u}) =\begin{bmatrix}
 -\mathcal{J}( \psi,\nabla^2 \psi) + \frac{1}{\mu_0\rho} \mathcal{J}( A, \nabla^2 A)\\
\mathcal{J}(\psi,  A)
 \end{bmatrix}
 \end{equation}

Consequently, the nonlinear terms in the equations above can be explicitly written in terms of the eigenmodes as:

\begin{equation}
\mathcal{J}(\psi_k,\nabla^2\psi_l)+\mathcal{J}(\psi_l,\nabla^2\psi_k)=\Bigg ( im_kP_{n_k}^{m_k}\frac{dP_{n_l}^{m_l}}{d\theta} -im_lP_{n_l}^{m_l}\frac{dP_{n_k}^{m_k}}{d\theta}\Bigg )\frac{\omega_k}{m_k}\frac{\omega_l}{m_l}\frac{1}{a^2cos\theta}\times\\
\bigg(n_l(n_l+1)-n_k(n_k+1))\bigg)e^{i(m_k+m_l)\phi}
\end{equation}

\begin{equation}
\frac{1}{\mu_0\rho}[\mathcal{J}(A_k,\nabla^2A_l) + \mathcal{J}(A_l,\nabla^2A_k)] =\Bigg ( im_kP_{n_k}^{m_k}\frac{dP_{n_l}^{m_l}}{d\theta} -im_lP_{n_l}^{m_l}\frac{dP_{n_k}^{m_k}}{d\theta}\Bigg)\frac{B_0^2}{\mu_0\rho a^2}\frac{1}{a^2cos\theta} \times \\
\bigg(n_l(n_l+1)-n_k(n_k+1))\bigg)e^{i(m_k+m_l)\phi}
\end{equation}

\begin{align}
\mathcal{J}(\psi_l,A_k) +\mathcal{J}(\psi_k,A_l)=
\Bigg ( im_kP_{n_k}^{m_k}\frac{dP_{n_l}^{m_l}}{d\theta} -im_lP_{n_l}^{m_l}\frac{dP_{n_k}^{m_k}}{d\theta}\Bigg )
\frac{B_0^2}{a^2}\Bigg(\frac{\omega_k}{m_k} -\frac{\omega_l}{m_l}\Bigg)\frac{1}{a^2cos\theta} e^{i(m_k+m_l)\phi}
\end{align}

Evaluating the integrals in (\ref{EqSpecAmplitGenericApendTriad}), as well as the inner product according to (\ref{DefInnerProd}), it follows that equation (\ref{EqSpecAmplitGenericApendTriad}) becomes

\begin{equation}
\frac{d\Lambda_j}{dt} - i\omega_j \Lambda_j = \Lambda_k(t) \Lambda_l(t) C_{j,k,l}
\end{equation}

with the coupling coefficient $C_{j,k,l}$ being expressed according to

\begin{equation}
C_{j,k,l}=
\frac{-i}{2a}K^{m_{1}m_{2}m_{3}}_{n_{1}n_{2}n_{3}}\bigg[
I^{m_{1}m_{2}m_{3}}_{n_{1}n_{2}n_{3}}+L^{m_{1}m_{2}m_{3}}_{n_{1}n_{2}n_{3}}\bigg]
\end{equation}

where the constants $I^{m_{j}m_{k}m_{l}}_{n_{j}n_{k}n_{l}}$, $L^{m_{j}m_{k}m_{l}}_{n_{j}n_{k}n_{l}}$ and the coupling integral $K^{m_{1}m_{2}m_{3}}_{n_{1}n_{2}n_{3}}$ are given by

\begin{equation}
I^{m_{j}m_{k}m_{l}}_{n_{j}n_{k}n_{l}}=
\Big [ (n_{k}(n_{k}+1))-(n_{l}(n_{l }+1))\Big ]
\end{equation}

\begin{equation}
L^{m_{j}m_{k}m_{l}}_{n_{j}n_{k}n_{l}}=\big(\frac{V_A}{a}\big)^2\Bigg( \frac{\omega_l}{m_l}-\frac{\omega_k}{m_k} \Bigg)
\end{equation}

\begin{equation}
K^{m_{j}m_{k}m_{l}}_{n_{j}n_{k}n_{l}}=N_{n_{j}}^{m_{j}}N_{n_{k}}^{m_{k}}N_{n_{l}}^{m_{l}}\int_{-1}^{1}P^{m{j}}_{n_{j}} \bigg( m_k P^{m_{k}}_{n_{k}}\frac{dP^{m_{l}}_{n_{l}}}{dz} -m_lP^{m_{l}}_{n_{l}}\frac{dP^{m_{k}}_{n_{k}}}{dz} \bigg)dz
\end{equation}

\section{The three-wave equations}\label{3WEqPRess}
Here we review some basic features on the dynamics of the three wave equations, including the precession resonance mechanism proposed by \cite{Bustamante2014}. Typically, in nonlinear wave problems with quadratic nonlinearities the nonlinear interactions involving the normal modes of the linear system are described by a complex chain of three-wave systems of the form: 

\begin{align} 
\frac{d\Lambda_{1}}{dt}=i\omega_{1}\Lambda_{1}+C_{1}\Lambda_{2} ^{*}\Lambda_{3} \label{DynamicalSystemAmp1}\\
\frac{d\Lambda_{2}}{dt}=i\omega_{2}\Lambda_{2}+C_{2}\Lambda_{1} ^{*}\Lambda_{3} \label{DynamicalSystemAmp2}\\
\frac{d\Lambda_{3}}{dt}=i\omega_{3}\Lambda_{3}+C_{3}\Lambda_{1} \Lambda_{2} \label{DynamicalSystemAmp3}
\end{align}
where $\Lambda_i$ denotes the complex valued amplitude of the i-th wave, $\omega_{i}$ is the corresponding eigenfrequency and $C_{i}$ the corresponding coupling coefficient. The dynamical system described above has three independent conserved quantities, namely the Hamiltonian

\begin{equation}
H=Im(\Lambda_1\Lambda_2\Lambda^{*}_3)
\end{equation}
as well as two quantities called Manley-Rowe relations:
\begin{align*} 
I_{12}=|\Lambda_1|^2+|\Lambda_2|^2\\
I_{13}=|\Lambda_1|^2+|\Lambda_3|^2
\end{align*}
These three conserved quantities make the system integrable. Using the polar representation $\Lambda_j = A_je^{i\Phi_j}$, the complex equations (\ref{DynamicalSystemAmp1})-(\ref{DynamicalSystemAmp3}) can be re-written as four equations describing the time evolution of the real amplitudes and the combination of the phases represented by $\Phi$:

 \begin{align} 
\frac{dA_{1}}{dt}=C_{1}A_{2} A_{3}cos\Phi\\
\frac{dA_{2}}{dt}=C_{2}A_{1} A_{3}cos\Phi\\
\frac{dA_{3}}{dt}=-C_{3}A_{1} A_{2}cos\Phi\\
\frac{d\Phi}{dt}=\Delta\omega + A_{1}A_{2}A_{3}\Bigg( \frac{C_{1}}{A_{1}^2}+\frac{C_{2}}{A_{2}^2}+\frac{C_{3}}{A_{3}^2}\Bigg) \\
\end{align}

where $\Delta\omega = \omega_3 - \omega_2 - \omega_1$ is the mismatch among the mode eigenfrequencies. With the conserved quantities described above, these equations are integrable by quadrature, with the solutions being expressed in terms of Jacobi elliptic functions (see \citealt{Bustamante2011} for details):

 \begin{align} 
A_{1}^2=-\mu\Big(\frac{2K(\mu)}{ZT}\Big)^2 sn^2 \Big(\frac{2K(\mu)(t-t_0)}{T},\mu\Big)+ \frac{I_{13}}{3}\Big( 2-\rho +2\sqrt(1-\rho-\rho^2)cos(\alpha/2)\Big)\\
A_{2}^2=-\mu\Big(\frac{2K(\mu)}{ZT}\Big)^2 sn^2 \Big(\frac{2K(\mu)(t-t_0)}{T},\mu\Big)+ \frac{I_{13}}{3}\Big( 2\rho -1+2\sqrt(1-\rho-\rho^2)cos(\alpha/2)\Big)\\
A_{3}^2=\mu\Big(\frac{2K(\mu)}{ZT}\Big)^2 sn^2 \Big(\frac{2K(\mu)(t-t_0)}{T},\mu\Big)+ \frac{I_{13}}{3}\Big( \rho+1-2\sqrt(1-\rho-\rho^2)cos(\alpha/2)\Big)\\
\end{align}

In the equations above, $sn$ stands for the elliptic sine, and $K(\mu)$ is the elliptic integral of the first kind, given by

\begin{equation}
K(\mu)=\int_{0}^{\pi/2}\frac{d\theta}{\sqrt{1-\mu sin \theta}}
\end{equation}

where the argument $\mu$ of the elliptic integral above is

\begin{equation}
\mu=\frac{cos(\alpha/3+\pi/6)}{cos(\alpha/3-\pi/6)}
\end{equation}

The solutions described above are periodic in time, with period T given by:

\begin{equation}\label{PeriodTriadSol}
T=\frac{\sqrt{2\sqrt{3}}K(\mu)}{Z(1-\rho-\rho^2)\sqrt{I_{13} \cos\big(\alpha/3-\pi/6}\big)}
\end{equation}

where $\rho$ is the ratio between two Manley-Rowe constants, $\rho=I_{13}/I_{23}$, and the angle $\alpha \in [0,\pi]$ is defined by

\begin{equation}
\cos\alpha =\frac{\sqrt{2\sqrt{3}}K(\mu)}{Z(1-\rho-\rho^2)\sqrt{I_{13}\cos\big(\alpha/3-\pi/6\big)}}
\end{equation}

Likewise, the solution for the phases is given by:

\begin{equation}
\Phi(t)=sign(\Phi_0)arccot\Bigg(\frac{\mu}{|H|}\Big(\frac{2K(\mu)}{ZT}\Big)^3  sn \Big(\frac{2K(\mu)(t-t_0)}{T},\mu\Big) cn \Big(\frac{2K(\mu)(t-t_0)}{T},\mu\Big) dn \Big(\frac{2K(\mu)(t-t_0)}{T},\mu\Big) \Bigg)
\end{equation}

with $dn$ and $cn$ indicating the other Jacobi elliptic functions.

\section{Two robust energy-transfer mechanisms in a five-wave model}\label{5WEqPRess}

Let us couple two triads of nonlinearly interacting waves through one mode, resulting in the five-wave model

\begin{align}
\frac{d\Lambda_{1}}{dt}=i\omega_{1}\Lambda_{1}+C_{123}\Lambda_{2} ^{*}\Lambda_{3} \label{AppFiveWaveEquationsa} \\
\frac{d\Lambda_{2}}{dt}=i\omega_{2}\Lambda_{2}+C_{231}\Lambda_{1} ^{*}\Lambda_{3} \label{AppFiveWaveEquationsb} \\
\frac{d\Lambda_{3}}{dt}=i\omega_{3}\Lambda_{3}+C_{312}\Lambda_{1} \Lambda_{2} + C_{345}\Lambda_{4}^{*} \Lambda_{5} 
\label{AppFiveWaveEquationsc} \\
\frac{d\Lambda_{4}}{dt}=i\omega_{4}\Lambda_{4}+C_{435}\Lambda_{5} ^{*}\Lambda_{3} \label{AppFiveWaveEquationsd} \\
\frac{d\Lambda_{5}}{dt}=i\omega_{5}\Lambda_{5}+C_{534}\Lambda_{3} \Lambda_{4} \label{AppFiveWaveEquationse}
\end{align}
where $\omega_j$ and $C_{ijk}$ can be read off  from Table \ref{tab:freqs}, namely: 
$$\omega_1=0,\,\omega_2=1.78236 \times 10^{-7},\,\omega_3=1.72695 \times 10^{-7},\,\omega_4=1.85859 \times 10^{-7},\,\omega_5=3.56473 \times 10^{-7},$$
and
$$C_{123} = -0.200293 i, \quad C_{231} = -1.75195 i, \quad C_{312} = -2.15463 i,$$
$$C_{345} = 0.620163 i, \quad C_{453} = 0.27978 i, \quad C_{534} = 0.904184 i.$$

 In general, equations (\ref{AppFiveWaveEquationsa})--(\ref{AppFiveWaveEquationse}) are not integrable, but it is easy to show that the following quadratic functions of the dependent variables $\{\Lambda_j\}_{j=1}^5$ are constants of the motion (Manley-Rowe invariants):
\begin{eqnarray}
\label{eq:Manley-RoweMB1}
I &=& |\Lambda_3|^2 +\frac{|C_{345}|}{|C_{534}|} |\Lambda_5|^2 +\frac{|C_{312}|}{|C_{123}|} |\Lambda_1|^2 ,\\
%I_{2} &=& |A_3|^2 +\frac{|C_{345}|}{|C_{534}|} |A_5|^2 +\frac{|C_{231}|}{|C_{123}|} |A_2|^2 , \\
\label{eq:Manley-RoweMB2}
J&=& \frac{|\Lambda_1|^2}{|C_{123}|}  - \frac{|\Lambda_2|^2}{|C_{231}|}  , \\
\label{eq:Manley-RoweMB3}
K &=& \frac{|\Lambda_4|^2}{|C_{453}|} + \frac{|\Lambda_5|^2}{|C_{534}|}\,.
\end{eqnarray}
The constancy of $J$ has a direct interpretation: $|\Lambda_1|^2$ and $|\Lambda_2|^2$ are directly coupled, as evidenced by the plots of the energies of modes with spherical wavenumbers $(0,2)$ and $(1,10)$ in figures \ref{Figure1} and \ref{Figure4}. Similarly, the constancy of $K$ means the direct coupling of $|\Lambda_4|^2$ and $|\Lambda_5|^2$, corresponding to spherical wavenumbers $(1,12)$ and $(2,10)$ in the same figures. Finally, the constancy of $I$  represents the coupling between triads \textbf{a} and \textbf{b}, and means that the energies $|A_1|^2, |A_3|^2, |A_5|^2$ must lie on a certain spheroid. Notably, from the fact that the numerical factors in these formulae take finite values it follows that all modes' energies are bounded from above.

\begin{figure}
  \includegraphics[width=\linewidth]{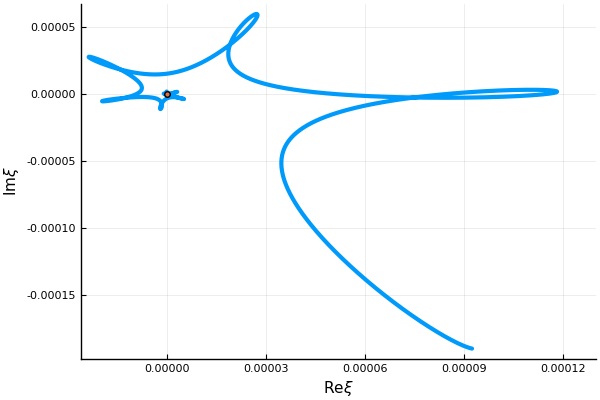}
  \caption{Real and imaginary parts of the mode 4  in the conservative  5-wave model linearized around the solution of the first triad in a 100 years integration. The growth of this mode is a result of a modulational type instability with a growth rate of .28/yr}
  \label{FigureA}
\end{figure}

Below we discuss briefly two robust mechanisms of energy transfer between the triads:
 
\noindent \textbf{Modulational Instability.} Note that  in case the amplitude and mismatch frequencies are commensurable  a particular solution of triad \textbf{b} alone defines a periodic orbit of the system of five waves. This is done by setting initial conditions $\Lambda_{1}=\Lambda_{2}=0$ at $t=0$ and the first and second modes' amplitudes will remain zero for all times. We can in principle linearize the system around a periodic solution of triad \textbf{b} and analyze the stability of the system to small perturbations on the first and second amplitudes. Such instability is reminiscent of the modulational instability  explored in \citealt{Connaughton2010} in the case of Rossby waves. In the periodic case this instability can be studied by using Floquet analysis (\citealt{Hale1969}). However, in general the triad equations are quasi-periodic so the instability analysis can be done by calculating the largest Lyapunov exponent of the system. In order to do this we use the procedure of \citealt{Benettin1976} and the implementation available in \citealt{Datseris2018}. In the case of system (\ref{AppFiveWaveEquationsa})--(\ref{AppFiveWaveEquationse}), with initial amplitudes for triad \textbf{b}: $\Lambda_3(0) = 9.56 (1+i)  \times 10^{-9}, \,\, \Lambda_4(0) = 10^{-2} \Lambda_3(0), \,\, \Lambda_5(0) = \frac{3}{2} \Lambda_3(0) $ (and infinitesimally small initial amplitudes for modes $\Lambda_1, \Lambda_2$), the largest Lyapunov exponent is $9 \times 10^{-9}$, corresponding to a growth rate of 0.28/yr which seems compatible with the time scales associated with the solar cycle. We illustrate the instability by plotting the real and imaginary parts of the mode $\Lambda_4$ in a 100-year integration in Figure  \ref{FigureA}.

%we have the solution of the three first components $(\Lambda_{1},\Lambda_{2}, \Lambda_{3})$, we can write the evolution equation for infinitesimal perturbations in the fourth and fifth equations as a linear system with periodic coeficients

%\begin{equation}
%\dot{X}=A(t)X
%\end{equation}
%with $X=(\Lambda_{4},\Lambda_{5})^T$, and
%\begin{equation}
%A(t)=
%\begin{bmatrix}
%i\omega_{4}&&C_{4,3,5}\Lambda_{3}\\
%C_{5,3,4}\Lambda_{3}&&i\omega_{5}
%\end{bmatrix}
%\end{equation}
%A periodic matrix with period T. If we consider the fundamental solution of this system $\Phi(t)$, which satisfies after a  period T,  $\Phi(t+T)=\Phi(t)\Phi^{-1}(0)\Phi(T)$, here $\Phi^{-1}(0)\Phi(T)$ is called the monodromy matrix. Calculation of the eigenvalues of this matrix, often called Floquet exponents determine the stability of the periodic solution under infinitesimal perturbations in the components $\Lambda_{4}$ and $\Lambda_{5}$.

\begin{figure}
  \includegraphics[width=\linewidth]{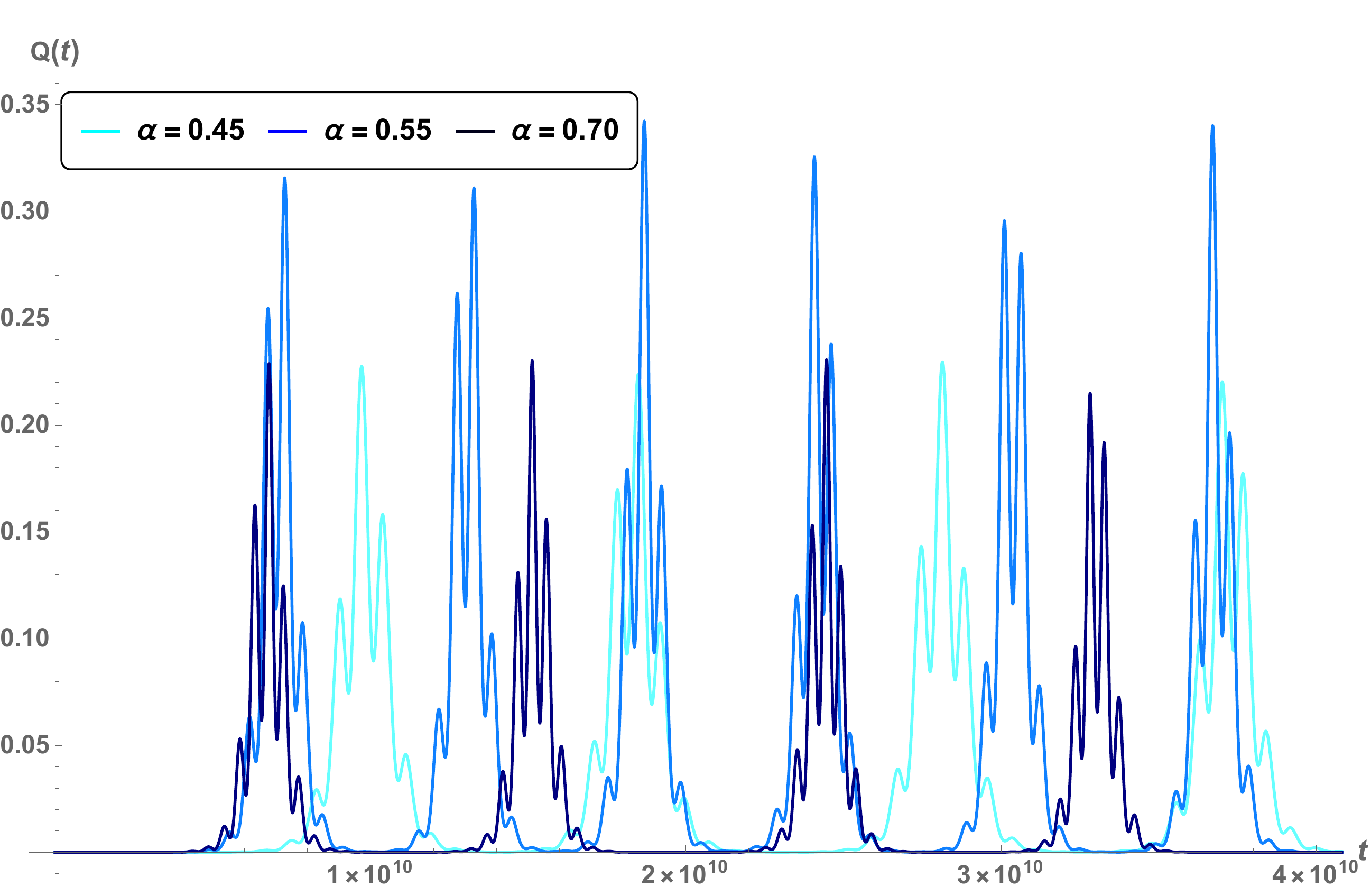}
  \caption{Evolution in time (over $1268$ years) of quotients ${\mathcal Q}(t)$ from equation (\ref{eq:Q(t)}) for different values of scale parameter: $\alpha=0.45$ (light blue), $\alpha = 0.55$ (blue), $\alpha = 0.70$ (dark blue). The maximum over time for each plot defines the efficiency $\mathcal{E}(\alpha)$, giving $\mathcal{E}(0.45) = 0.23, \,\,\mathcal{E}(0.55) = 0.34, \,\,\mathcal{E}(0.70) = 0.23$.}
  \label{FigureQ}
\end{figure}

\begin{figure}
  \includegraphics[width=\linewidth]{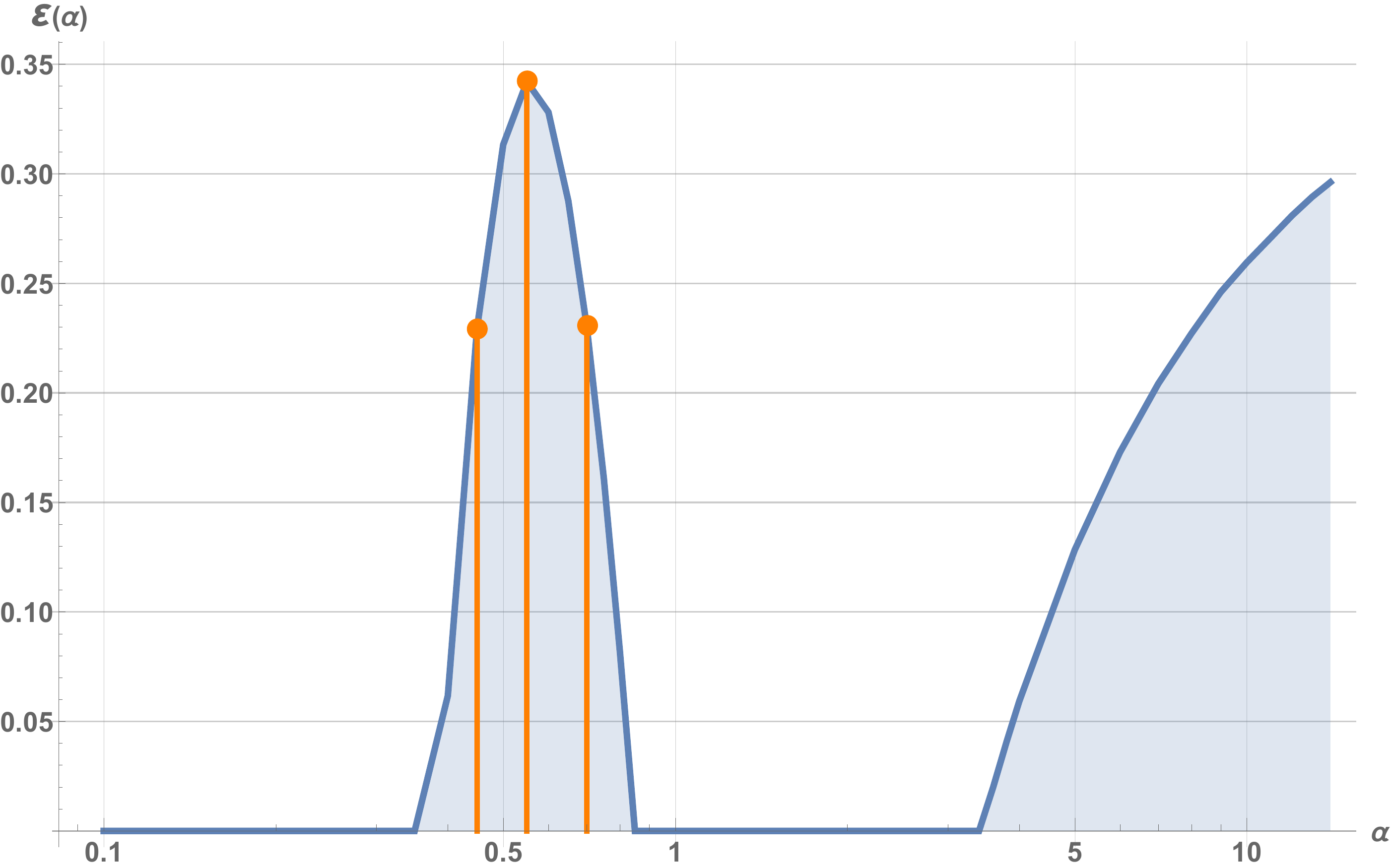}
  \caption{Solid curve: efficiency $\mathcal{E}(\alpha)$, calculated over a time evolution of $1524$ years, for a wide range of the scale parameter: $\alpha \in [0.1, 14.0]$. This range is shown in $\log$ scale to aid visualisation. The highest efficiency ($34\%$) is attained at $\alpha = 0.55$, significantly higher than the efficiencies attained at large amplitudes $\alpha \gg 1$. Vertical markers: selected values of $\alpha = (0.45, 0.55, 0.70)$ used for figure \ref{FigureQ}.}
  \label{FigureEff}
\end{figure}

\noindent \textbf{Precession resonance.} To explain this mechanism let us consider a simple instance whereby triad (1, 2, 3) has low energy in comparison with triad (3, 4, 5) initially. Let us also assume, for simplicity of exposition, that triad (3, 4, 5) is resonant or quasi-resonant ($\Delta\omega_2 \equiv \omega_3+\omega_4-\omega_5 \approx 0$). In this case, as presented before, the time evolution of the mode amplitudes of triad (3, 4, 5) is periodic in time, with period $T_2$ inversely proportional to the wave amplitudes according to (\ref{PeriodTriadSol}). Consequently, in this appropriate amplitude regime in which the $\Delta\omega_1$ term dominates the corresponding time evolution equation of the relative phase of triad (1, 2, 3), if $ \frac{2\pi}{T_2} \approx \Delta\omega_1$, the last term in the RHS of equation (\ref{AppFiveWaveEquationsc}) will act as a resonant forcing for triad (1, 2, 3), making the energy of this triad to grow in time. This resonance between the linear frequency mismatch of one triad and the nonlinear frequency of energy oscillation of the other one was called precession resonance by \cite{Bustamante2014}. In this resonant case, \cite{Bustamante2014} showed that there is a strong energy transfer between different wave triads even in the case of non-resonant interactions ($\Delta\omega_1 \neq 0$). To illustrate this effect, we consider a more appropriate set of initial conditions: $\Lambda_1 = 1.48 \alpha \exp(1.53 i) \times 10^{-13} , \Lambda_2 = 1.71 \alpha \exp(0.304 i) \times 10^{-13}, \Lambda_3 = 1.03 \alpha \exp(1.11 i)  \times 10^{-8}, \Lambda_4 = 1.22 \alpha  \exp(5.06 i)\times 10^{-8}, 
\Lambda_5 = 1.17 \alpha  \exp(3.95 i)\times 10^{-8}$, where $\alpha$ is a real scale parameter (of order one) used to search for the resonance. The phases of these initial conditions were randomly generated uniformly over $[0,2\pi]$;   apart from the scale parameter $\alpha$, the amplitudes $|\Lambda_1|, |\Lambda_2|$ were randomly generated uniformly over the domain $[1, 2]\times 10^{-13}$ and the amplitudes $|\Lambda_3|, |\Lambda_4|, |\Lambda_5|$ were randomly generated uniformly over the domain $[1,2]\times 10^{-8}$. In order to find the resonance, a simulation of system (\ref{AppFiveWaveEquationsa})--(\ref{AppFiveWaveEquationse}) is done for selected choices of scale parameter $\alpha$. For each simulation, the quotient 
\begin{equation}
\label{eq:Q(t)}
\mathcal{Q}(t) \equiv \frac{|C_{312}| \,|\Lambda_1(t)|^2}{|C_{123}|\, I}, \qquad  0 \leq \mathcal{Q}(t) \leq 1 \quad \text{for all}\quad t\geq 0,
\end{equation}
where $I$ is defined in equation (\ref{eq:Manley-RoweMB1}),
is plotted over a long time range (about $1524$ years in this work) and its maximum value over that time range is recorded, giving the so-called efficiency $\mathcal{E}(\alpha) \equiv \max_{t}\mathcal{Q}(t)$. Plots of $\mathcal{Q}(t)$ for selected values of scale parameter ($\alpha=0.45, 0.55$ and $0.70$) are shown in figure \ref{FigureQ} and a plot of efficiency $\mathcal{E}(\alpha)$ over an extended range $0.1 \leq \alpha \leq 14.0$ is shown in figure \ref{FigureEff}. Remarkably, the peak of efficiency at $\alpha=0.55$ gives $34\%$ efficiency, significantly larger than the calculated efficiency at unfeasibly higher amplitudes $\alpha \gg 1$. The time series plotted in figure \ref{FigureQ} show three time scales: a fast one corresponding to the typical nonlinear time scale of the order of 10 years, an intermediate one corresponding to the envelopes' widths, of the order of 100 years, and a slow one corresponding to the distance between the envelopes, which can vary between 100 and 1000 years as we could measure in our extended sweep over values of $\alpha$ between $0.1$ and $14$. The energy share of this long-time range, as a function of $\alpha$, has a marked peak in the vicinity of the peak at $\alpha=0.55$ and also in the vicinity of the transition point at $\alpha = 3.40$ (figure not shown). These low-frequency peaks provide evidence that precession resonance is the mechanism behind the observed strong energy transfers toward zonal modes.

\section{Evaluating the Damping Coefficients}\label{ApendDifus}

Due to the fact that the dissipation coefficients are different for the velocity and magnetic fields, we have to project the resulting effects onto the eigenvectors corresponding to each wave mode in order to obtain the dissipation coefficients. Consider an equation for a freely decaying vector field:

\begin{equation}
\frac{\partial}{\partial t}\overrightarrow{V} = \mathcal{D}\overrightarrow{V}
\end{equation}

where
\begin{equation}
\overrightarrow{V}=\begin{bmatrix}\nabla^2\psi\\ A\end{bmatrix}
\end{equation}

and the linear dissipation operator is given by

\begin{equation}
\mathcal{D}=
\begin{bmatrix}
   \nu\nabla^2(\nabla^2) & 0\\
     0& \eta  \nabla^2(\nabla^2) 
  \end{bmatrix}
\end{equation}

As described before, in the equations above $\nu$  and $\eta$ are the coefficients of viscous and magnetic diffusivity, respectively.    In order to obtain the spectral amplitude equations associated with the freely decaying vector field, one multiplies the first component equation by $\psi$  and the second by $\nabla^2A$ and integrates by parts, obtaining in the spectral space the following equations for the amplitudes $\Lambda_j$:

 \begin{equation}
\frac{d \Lambda_j}{dt}= \sum_i \mathcal{D}_{ij}\Lambda_i
\end{equation}

where i, j denote the particular eigenmode, characterized by a spherical harmonic (m, n) and one of the mode types (slow magnetic or fast hydrodynamic branch). Because of the orthogonality relations of the spherical harmonics, a given mode can be influenced only by itself and by the corresponding mode with the same wavenumber but in the opposite branch (other type of wave), since the nonlinear interaction of one mode with another one with same wavenumber if forbidden by the Ellsaesser rules (\citealt{Ellsaesser1966}). Therefore, we consider only diagonal interactions, whose coefficients are given by (\ref{FinalExpressCoefDifus}).

\startlongtable
\begin{deluxetable}{c|ccc}
\tablecaption{ \label{tab:freqs} Wave-numbers and corresponding eigenfrequencies and coupling coefficients of the selected waves in the five wave-model, separated in two triads (a and b). The corresponding frequency mismatches give $1/\Delta \omega_a=5.72248 yrs$ and $1/\Delta \omega_b=15.2326 yrs$.}
\tablehead{
\colhead{Wavenumber} & \colhead{Eigenfrequency(Hz)}  & \colhead{Triad} &\colhead{Coupling  coefficient} \\
}
%\colnumbers
\startdata
(0, 2) & $0$ &a &- 0.200293i\\
(1, 10)  & $1.78236*10^{-7}$&a  & - 1.75195i\\
(1, 9)  &  $1.72695*10^{-7 }$ &a&- 2.15463i\\
(1, 9)&  $1.72695*10^{-7}$&b  & 0.620163i\\
(1, 12) &$1.85859*10^{-7} $ &b & 0.27978i\\
(2, 10) & $3.56473*10^{-7} $&b &0.904184i\\
\enddata
%\tablecomments{}
\end{deluxetable}

\end{document}